\documentclass[prl,aps,twocolumn,floats,floatfix]{revtex4}
\usepackage{graphicx}  % Include figure files
\usepackage{dcolumn}   % Align table columns on decimal point
\usepackage{amsmath}
\usepackage{amssymb}
\usepackage{subfigure}
\usepackage{color}
\usepackage{bm}% bold math
\usepackage{natbib}% bold math

\begin{document}
\bibliographystyle{apsrev}
\title{\bf On the Impact of Solvation on 
a Au/TiO$_2$ Nanocatalyst in Contact with Water}
%dm \title{\bf Unraveling Solvent Effects at a Au/TiO$_2$(110)-Water Interface Nanocatalyst }
\author{Matteo Farnesi Camellone} 
\affiliation{Lehrstuhl f\"ur Theoretische Chemie, Ruhr--Universit\"at Bochum,
44780 Bochum, Germany}

\author{Dominik Marx} 
\affiliation{Lehrstuhl f\"ur Theoretische Chemie, Ruhr--Universit\"at Bochum,
44780 Bochum, Germany}

\date{\today}% It is always \today, today,
%\abstract{%
\begin{abstract}
Water, the ubiquitous solvent, is also prominent in
forming liquid-solid interfaces with catalytically active surfaces,
in particular with promoted oxides. 
We study the complex interface of a gold nanocatalyst, 
pinned by an F--center on titania support,  
and water.
The {\em ab initio} simulations uncover 
%surprising
%MF removed the word "surprising" as requested by referee 1 
the microscopic 
details of solvent-induced charge rearrangements at the metal particle.
Water is found to stabilize charge states 
differently from the gas phase as a result of
structure-specific charge transfer from/to the solvent, thus 
altering  
surface reactivity. 
The metal cluster is shown to feature both ``cationic''
and ``anionic'' solvation, depending on fluctuation and 
polarization effects in the liquid,
which creates novel active sites. 
These observations open up an avenue toward ``solvent engineering''
in liquid-phase heterogeneous catalysis.\\ 
\end{abstract}
%}
%}

\maketitle

Highly dispersed gold nanoparticles supported on
oxides have been shown to catalyze a number of important reactions,
including low--temperature $\mathrm{CO}$ oxidation and the water gas shift
reaction~\cite{haruta,boccuzzi}.            
Reducible oxides, in particular titania ($\mathrm{TiO_{2}}$), 
are ideal catalytic supports~\cite{diebold}.
The size of the gold particles substantially affects the catalytic activity,
suggesting the key importance of metal/support 
interactions on a nanometer scale~\cite{valden}. 
Reactions, and in particular $\mathrm{CO}$ oxidation, are believed 
to occur at specific active $\mathrm{Au}$ sites at the
$\mathrm{Au}$/$\mathrm{TiO_{2}}$ interface~\cite{liu1,hammer,yates,behm}. 
Although much is known regarding $\mathrm{Au}$/$\mathrm{TiO_{2}}$ catalytic
activity in the presence of a gas phase,
the complexity increases steeply
when solvent is included. 

Liquid--solid interfaces as such are relevant to many industrial applications 
of great significance, such as \mbox{(photo-)catalysis}, 
solar cells,
gas sensors, or biocompatible devices.
In heterogeneous catalysis, it has been shown that
the presence of water  
increases the observed rate of CO oxidation~\cite{rao,date}.
%
%
%added text to improve the introduction explaing better the ability of Au/TiO2-H2O
%to convert CO in CO2
%
%
The degree of rate enhancement depends on the type of support used. 
In particular for the case of Au/TiO$_{2}$ catalysts 
it has been shown that 
%dm in the case of Au/TiO$_{2}$ its 
their 
activity at about 3000~ppm H$_{2}$O is so high that 
full conversion of CO is reached~\cite{date}. 
Thus, 
the Au/TiO$_{2}$ surface displays a pronounced catalytic activity toward 
the water-gas-shift 
(WGS)
%dm (WGSR)
reaction. 
This fundamental reaction represents a key reaction to produce extra H$_{2}$
fuel from steam reforming, which is reversible and exothermic, according to
the following reaction: CO + H$_{2}$O 
$\leftrightharpoons$ 
%dm $\leftrightarrow$ 
CO$_{2}$ + H$_{2}$. 
The so--called carboxyl and redox mechanisms have been proposed 
for the WGS reaction on metal/oxide surfaces; in the former 
%dm (carboxyl) 
mechanism the CO species reacts with
terminal hydroxyl groups, whereas in the latter CO reacts with an O~atom
from OH dissociation or from the oxide support. 
In both mechanism proposed the starting point is a 
H$_{2}$O molecule that is initially adsorbed on the metallic
cluster with its oxygen atom being attached to a metal atom. 
%dm cluster with the oxygen atom of the H$_{2}$O attached to a metal atom. }

%
%dm Even
Moreover, even 
large (as opposed to nanoscale) gold particles, 
which are usually 
catalytically inert, show considerable oxidation activity 
at aqueous conditions~\cite{sanch,zhi}.
Because of its high dielectric constant, water
may actively participate in chemical reactions for instance
by stabilizing ionic species,  
thus speeding up reactions at liquid-solid interfaces,
which opens up novel avenues to improve the performance 
of traditional heterogeneous catalysis at the gas-solid interface. 
Recently, it has been shown that the selective oxidation of
alcohols in aqueous phase over ($\mathrm{TiO_{2}}$, $\mathrm{C}$)--supported
$\mathrm{Au}$ catalysts is facilitated by high~pH conditions~\cite{davis}.
This  
water-based approach to heterogeneous catalysis 
offers a sustainable, environmentally
benign, and cheap  
alternative to traditional processes that rely on 
both toxic and expensive inorganic oxidants and organic solvents~\cite{corma}.
Despite these promising experimental findings, much remains unknown 
at the molecular level about the impact of water as a solvent 
on the reactivity of these modern catalyst systems.

Several {\it ab initio} molecular dynamics (AIMD)~\cite{marx-hutter-book} 
studies have been recently reported focussing
on the 
fundamentals of the
water-titania interface.
For a water film on $\mathrm{TiO_{2}}(110)$ 
it
has been shown that
at least two distinct layers form:
molecules in the first layer 
are sluggish and bind strongly to fivefold coordinated
$\mathrm{Ti}$ sites, whereas those in the second layer interact
only weakly with the substrate and diffuse rapidly~\cite{liu}, 
thus yielding a highly anisotropic interface~\cite{jphyslett}.
Moreover, 
water was not seen to dissociate at the coverages examined.    
Upon studying the thermodynamics of de/protonation 
of the 
rutile-water interface using free energy perturbation methods,
a
value of 0.6~eV was found for the dissociation free energy of 
bulk water on defect-free rutile~\cite{sprik}.
These studies thus show that the ideal  
$\mathrm{TiO_{2}}$ surface is
rather inert with respect to solvation by water. 
However, little is known
about the interaction between water and
defective or metal-supporting $\mathrm{TiO_{2}}$ interfaces,  
being relevant to catalysis and industrial applications. 
%

\begin{figure}[ht]
\begin{center}
\begin{tabular}{cc}
\includegraphics[angle=0,scale=0.075]{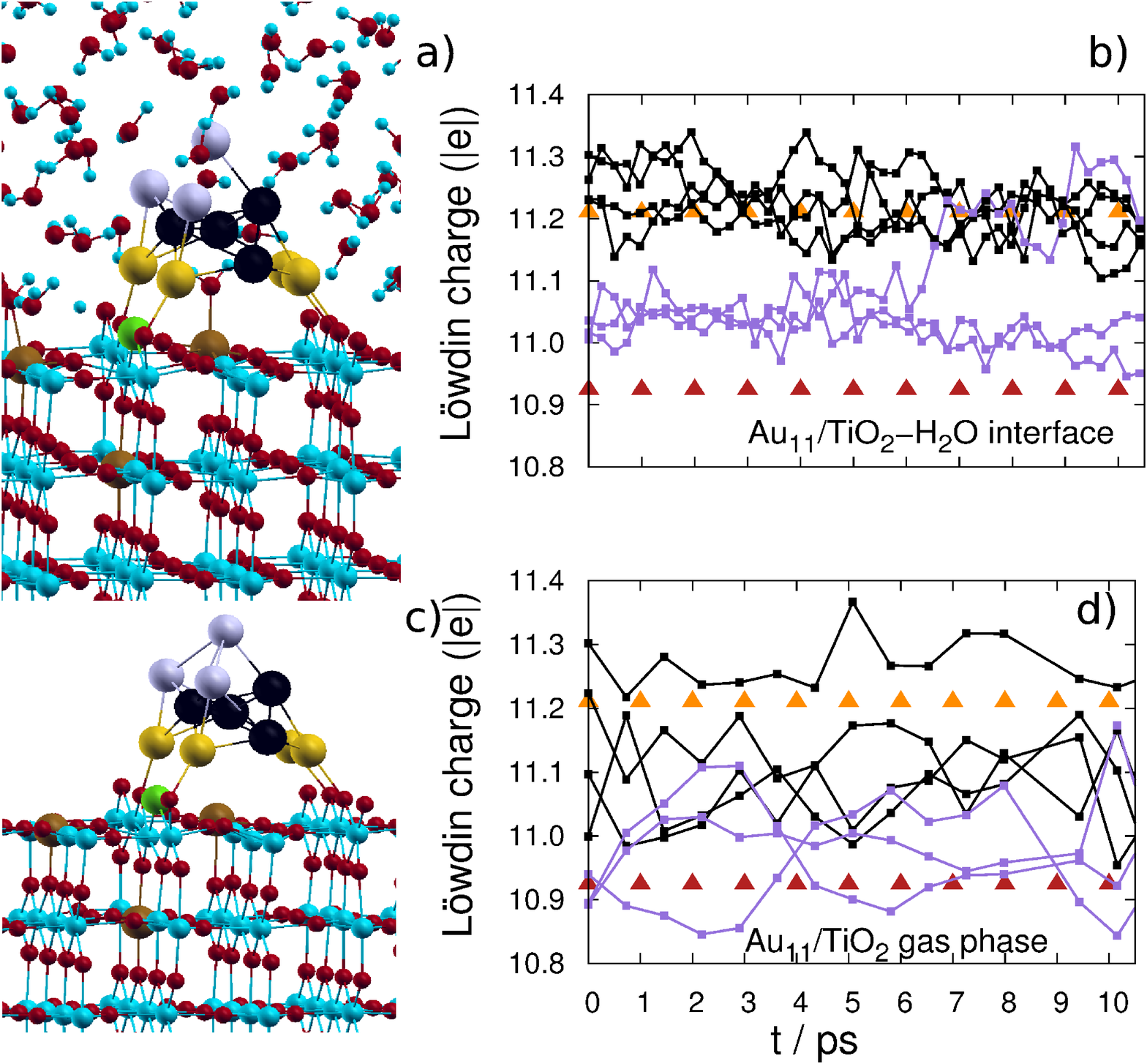}
\end{tabular}
\caption{
\footnotesize
{\bf Ball and stick model of the Au$_{11}$ nanocluster pinned by an 
F--center on the TiO$_{2}$(110) oxide surface in contact 
with  
liquid water~(a)  
and with a gas phase~(c).} 
Red, big blue, small blue and brown spheres are 
$\mathrm{O}$, $\mathrm{Ti}$, $\mathrm{H}$  
and $\mathrm{Ti^{3+}}$ sites, respectively, whereas the 
surface $\mathrm{O}$~vacancy is shown in green. 
Violet, black and yellow spheres correspond to top, middle and bottom
$\mathrm{Au}$ sites of the $\mathrm{Au_{11}}$ nanocluster, respectively.  
Time evolution of the L\"owdin charges of the top (violet lines) and 
middle (black lines) $\mathrm{Au}$ atoms of the supported gold nanocluster 
in liquid water (b) and in the gas phase (d). 
Red and orange triangles correspond to reference L\"owdin charges computed 
for a single Au$^{+}$(aq) cation (10.924~$|e|$) and Au$^{0}$(aq) atom (11.210~$|e|$)   
in liquid water, respectively. 
\label{fig1}}
\end{center}
\end{figure}

%
%In this Letter, we describe
% mf1
Here, we present large-scale 
AIMD simulations aimed to investigate solvent effects at
a gold nano\-cluster pinned by an 
F--center on TiO$_{2}$(110) 
being in contact with liquid water.  
The aqueous solution is found to induce pronounced charge 
transfer and localization at the nanocatalyst--liquid interface
and stabilizes structures different from the gas phase.
The $\mathrm{TiO_{2}(110)}$ surface has been modeled by four
$\mathrm{O}$--$\mathrm{Ti}_2\mathrm{O}_2$--$\mathrm{O}$ tri--layer (4$\mathrm{\times}$2)
supercell slabs separated by more than 15~{\AA}.
The most common point defects on the TiO$_{2}$(110) rutile surface are oxygen
vacancies in the twofold coordinated $\mathrm{O}$ rows~\cite{diebold}.
Therefore, 
a $\mathrm{Au}_{11}$ nanocluster was grown on 
TiO$_{2}$(110) where an F--center
created by a surface $\mathrm{O}$~vacancy acted as anchoring site 
for initial $\mathrm{Au}$ nucleation;
note that this blocks the O~defect to interact with water and thus
prevents splitting of water molecules at this F--center. 
The lowest energy structure of $\mathrm{Au_{11}}$ adsorbed on a reduced rutile
surface and employed in the present study is shown in Fig.~\ref{fig1}(c)).
In order to create the
Au$_{11}$/TiO$_{2}$--water interface, the space between the 
slabs has been fully filled with 53 H$_{2}$O molecules 
(see Fig.~\ref{fig1}(a)).
It
has been shown that $\mathrm{Au_{7}}$ is the smallest Au cluster on rutile
$\mathrm{TiO_{2}(110)}$ with a measurable reaction rate for CO combustion~\cite{combustion}.
The $\mathrm{Au_{11}}$ model, although being a simplification of the nanometersized
clusters on $\mathrm{TiO_{2}(110)}$, appropriately mimics the active sites
located at the nanogold/oxide interface where the oxidation process
takes place~\cite{hammer,models1}.
%
%
%
%mf1 moved to the section methods 
%
%
%All calculations have been performed using spin--polarized $\mathrm{PBE+U}$~\cite{pbe},
%and ultrasoft pseudopotentials~\cite{vanderbilt} 
%as implemented in {\tt CPMD}~\cite{cpmd} 
%and {\tt Quantum} {\tt Espresso}~\cite{QE}.
%
%In line with our previous work~\cite{cite-also-our-basic-PRB-2009,PRL,PRB_ma}, 
%the value of $\mathrm{U}$ = 4.2~eV was adopted.
%
%The AIMD simulations~\cite{marx-hutter-book} 
%used Car--Parrinello propagation, 
%the canonical ensemble was established at 450~K
%with a Nos$\mathrm{\acute{e}}$-Hoover 
%chain
%thermostat, and 
%about $\sim 10$~ps trajectories have been generated for analysis
%(see SOM for details).
%
The binding of the Au$_{11}$ cluster on the reduced $\mathrm{TiO_{2}}$(110)
support entails a strong charge rearrangement at the metal/oxide
contact (see Supporting Figure 1), the adsorption energy being $-2.19$~eV. 
In the case of an isolated $\mathrm{O}$~vacancy on the stoichiometric
TiO$_{2}$(110) surface, the charge neutrality is maintained by the presence of
two Ti$^{3+}$ ions 
thus creating an F--center.
Spin density and bonding charge analyses reveal that the $\mathrm{Au_{11}}$
cluster, once adsorbed, leaves a reduced substrate with three Ti$^{3+}$ ions, 
and $\sim 0.4$~$|e|$ are transferred from the metal cluster to the oxide.
This results into a slightly positively charged
Au$_{11}^{\delta +}$ cluster supported 
by a reduced TiO$_{2}$(110) oxide surface. 

In order to reveal the solvent-induced charge rearrangement at the metal-liquid
contact, we have performed $\mathrm{PBE+U}$ AIMD simulations of
the Au$_{11}$/TiO$_{2}(110)$ nanocatalyst in aqueous solution.
Additionally, corresponding gas phase simulations have been carried out
to provide the solvent-free reference situation.  
In both cases the finite temperature dynamics 
preserves 
three reduced $\mathrm{Ti^{3+}}$ sites (see Supporting Figure 2).
%dm  along the full trajectories.
%
Snapshots of the AIMD simulations have been collected
every $\sim 0.2$~ps 
thus generating a set of representative configurations
for electronic structure analyses. 
In the following, we will refer to $\mathrm{Au}$~atoms of the
supported  $\mathrm{Au_{11}}$ nanocluster 
as top, middle and bottom sites, 
which are depicted as
violet, black and yellow spheres in Fig.~\ref{fig1}, 
respectively.
A detailed investigation of the interaction between $\mathrm{H_{2}O}$
molecules and the bare oxide surface far from the supported metal cluster is
out of the focus
of this 
%Letter
% mf1
study
, but
we note in passing that
our AIMD simulations are in agreement with those of~\cite{liu,jphyslett,sprik}
in the regions.
In particular, water at the liquid--oxide contact is slow,
strongly bonded to fivefold coordinated $\mathrm{Ti}$ sites,
and is not seen to dissociate.

The charge dynamics of the
$\mathrm{Au}$~sites 
was extracted by computing, as a function of time, 
the L\"owdin charges, 
see
Fig.~\ref{fig1}, where violet and black lines correspond 
to the charge evolution of top and middle $\mathrm{Au}$
atoms, respectively.   
As demonstrated by Fig.~\ref{fig1}(b), two distinct
charge patterns, corresponding to top and middle $\mathrm{Au}$~sites, 
can be clearly distinguished in aqueous solution;
note that $\mathrm{Au}$ atoms in direct contact with
TiO$_2$ (yellow spheres in Fig.~\ref{fig1}(a))
cannot be
solvated by the liquid phase and thus do not exhibit a clear trend.  
The average charges on top and middle $\mathrm{Au}$~atoms are
%DM The average charge on top and middle $\mathrm{Au}$~atoms are
11.028~$\pm$~0.006
and 11.220~$\pm$~0.006~$|e|$, 
respectively 
(see Supporting Table 1 for statistical analysis).
%
%DM
It is noted in passing that a very similar charge separation 
has been found for a gold cluster of different
size and shape using an Au$_{13}$/TiO$_{2}(110)$ nanocatalyst 
in contact with water.
Interestingly, at around 6~ps
a considerable amount of charge is transferred to a 
specific top $\mathrm{Au}$~site due to solvent fluctuations,
which reaches
a value that is typical of middle $\mathrm{Au}$ atoms,
see Fig.~\ref{fig1}(b).
In stark contrast, no such charge 
separation and fluctuation effects
can be identified in 
the gas phase reference dynamics in Fig.~\ref{fig1}(d). 
We conclude 
that the liquid phase promotes and stabilizes a net 
charge separation in the supported $\mathrm{Au}$ cluster, 
where top and middle atoms can be clearly distinguished
on the basis of their electronic structure.  
As a result, water steers the $\mathrm{Au}$ sites towards
either one or the other preferred charge value, depending on their 
specific location within the nanocluster and going hand in hand
with different solvation patterns, {\em vide infra}.

\begin{figure}
\begin{center}
\begin{tabular}{cc}
\includegraphics[angle=0,scale=0.39]{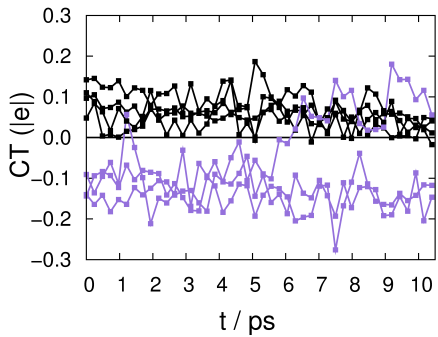}
\end{tabular}
\caption{%
\footnotesize 
{\bf Dynamics of the 
water~$\leftrightharpoons$~nanocluster 
charge transfer (CT).}
Violet and black lines correspond to top and middle $\mathrm{Au}$ atoms, respectively
%of top (violet lines) and middle
%(black lines) $\mathrm{Au}$ atoms.
 \label{fig2}}
\end{center}
\end{figure}

To further characterize this puzzling process we have quantified the 
amount of charge transfer (CT) from $\mathrm{Au}$~atoms 
to the aqueous solution 
and back  
using Bader analysis~\cite{bader,bader1}. 
For this purpose we have computed the difference between
the Bader charges of the $\mathrm{Au}$ atoms in the 
fully solvated Au$_{11}$/TiO$_{2}$(110)--water 
system and those of the corresponding 
Au$_{11}$/TiO$_{2}$(110) gas phase system, where the latter
is obtained upon removing the solvent while keeping all
atomic positions fixed.
As depicted in Fig.~\ref{fig2} we find that top $\mathrm{Au}$~sites 
transfer charge to the solvent, with 
an average CT of about 0.13~$\pm$~0.05~$|e|$,
while middle $\mathrm{Au}$~atoms 
attract 
about 0.06~$\pm$~0.03~$|e|$ per Au~atom from water.

In view of these pronounced CT effects, we probe 
solvent-induced morphology changes of the metal cluster 
with respect to the gas phase.  
A set of trajectory configurations  
of the solvated 
nanocatalyst has been selected
where water has been removed
before quenching and optimizing the remaining Au$_{11}$/TiO$_{2}(110)$ system. 
In order to generate the proper gas phase reference, 
Au$_{11}$/TiO$_{2}(110)$ in the absence of water
has been run at the same temperature of 450~K before 
applying the same quenching protocol;
%dm  . 
note that this is the relevant experimental temperature 
used for selective liquid--phase alcohol oxidation in
aqueous solution using gold/titania catalysts. 
In case of the 
quenched solvated interface,  
the same local minima and no significant change in the L\"owdin charges 
were observed
as in the solvated state. 
However,  
the gas phase reference system yielded 
several local minima 
with significant changes of both the real-space structure and the Au~charges 
(see Supporting Figures 3 and 4). 
These findings show that the liquid phase stabilizes
structures and charge states within nanogold that are 
distinctly different from those observed in
the gas phase, which eventually induces novel 
morphologies and active sites, respectively, due to solvation.

Let us 
now investigate the solvent-induced electronic charge redistribution 
at the nanocatalyst.
In the following, we refer the electronic
charge density of the solvated
Au$_{11}$/TiO$_{2}$(110)--water system 
to that of the isolated one,  Au$_{11}$/TiO$_{2}$(110), 
and the separated solvation water. 
We thus compute the difference, $\Delta \rho(\vec{r})$, 
between the charge densities of the combined system and the
separated components 
while freezing all atomic positions.
\begin{figure}
\begin{center}
\begin{tabular}{cc}
\includegraphics[angle=0,scale=0.4]{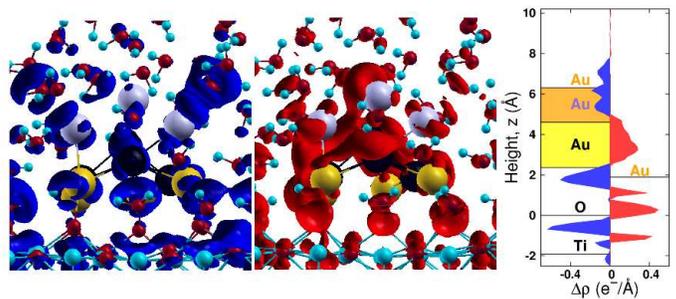}
\end{tabular}
\caption{%
\footnotesize 
{\bf Water-induced charge flow.}
Left and central panels: 
charge difference, 
$\Delta \rho(\vec{r})$, for a representative
snapshot of the Au$_{11}$/TiO$_{2}$(110)--water interface. 
Electron depletion/accumulation is depicted by blue/red 
isosurfaces 
at 
$\mathrm{\pm} 0.02$~$|e|$/$\mathrm{\AA ^{3}}$.
Right panel:  
charge difference
integrated in planes perpendicular to the surface, $\Delta \rho(z)$, 
as a function of the height, $z$.  
   \label{fig3}}
\end{center}
\end{figure}
A representative example of water-induced charge flow  
is depicted in Fig~\ref{fig3},
where the positive (negative) component of the charge difference 
$\Delta \rho_{+(-)}(\vec{r})$
is shown by red (blue) isosurfaces.  
This reveals 
that charge depletion (blue)
occurs mainly on top $\mathrm{Au}$ atoms, 
whereas additional electron charge (red)
prefers to accumulate on middle $\mathrm{Au}$ sites. 
Furthermore, 
charge depletion and accumulation visible 
at the oxide-water contact  
can be traced back to
the interaction between water and the
TiO$_{2}$(110) oxide support. 
Here charge depletion
occurs mainly on top of surface O~atoms 
to which $\mathrm{H_{2}O}$ molecules point with their $\mathrm{H}$ atoms, 
whereas charge accumulates 
close to
five-fold coordinated $\mathrm{Ti}$ atoms, being in turn coordinated by water O~atoms. 
This is quantified by considering 
the charge redistribution perpendicular to the support, 
$\Delta \rho(z)$
(see Fig.~\ref{fig3}). 
Here $\Delta \rho(z)$ is the sum
of the positive and negative  
charge density components 
and integrated in planes perpendicular to the surface.
This quantitative analysis reveals that, on the one hand, there is a net
charge transfer of $\sim 0.22$~$|e|$ from top $\mathrm{Au}$ atoms to
$\mathrm{H_{2}O}$ molecules (obtained by integrating $\Delta \rho(z)$ 
within the shaded orange area in Fig.~\ref{fig3}), leading
to slightly positively charged $\mathrm{Au^{\delta +}}$ atoms
at the corresponding sites. 
On the other hand, there is a reverse charge flow
of $\sim 0.37$~$|e|$ from water to middle $\mathrm{Au}$ atoms 
(see yellow area in
Fig.~\ref{fig3}), thus creating partially
negatively charged $\mathrm{Au^{\delta -}}$ atoms upon solvation. 
Similar results were obtained for all the representative structures 
sampled along the trajectory.
This can be compared to 
the average negative and positive 
CT obtained from the Bader analysis.
In total,
this amounts to a structure-specific charge exchange between
metal and water of $-0.34$ and $+0.46$~$|e|$, in qualitative agreement with
the charge difference analysis. 
\begin{figure}
\begin{center}
\begin{tabular}{cc}
\subfigure[]{\includegraphics[angle=0,scale=0.15]{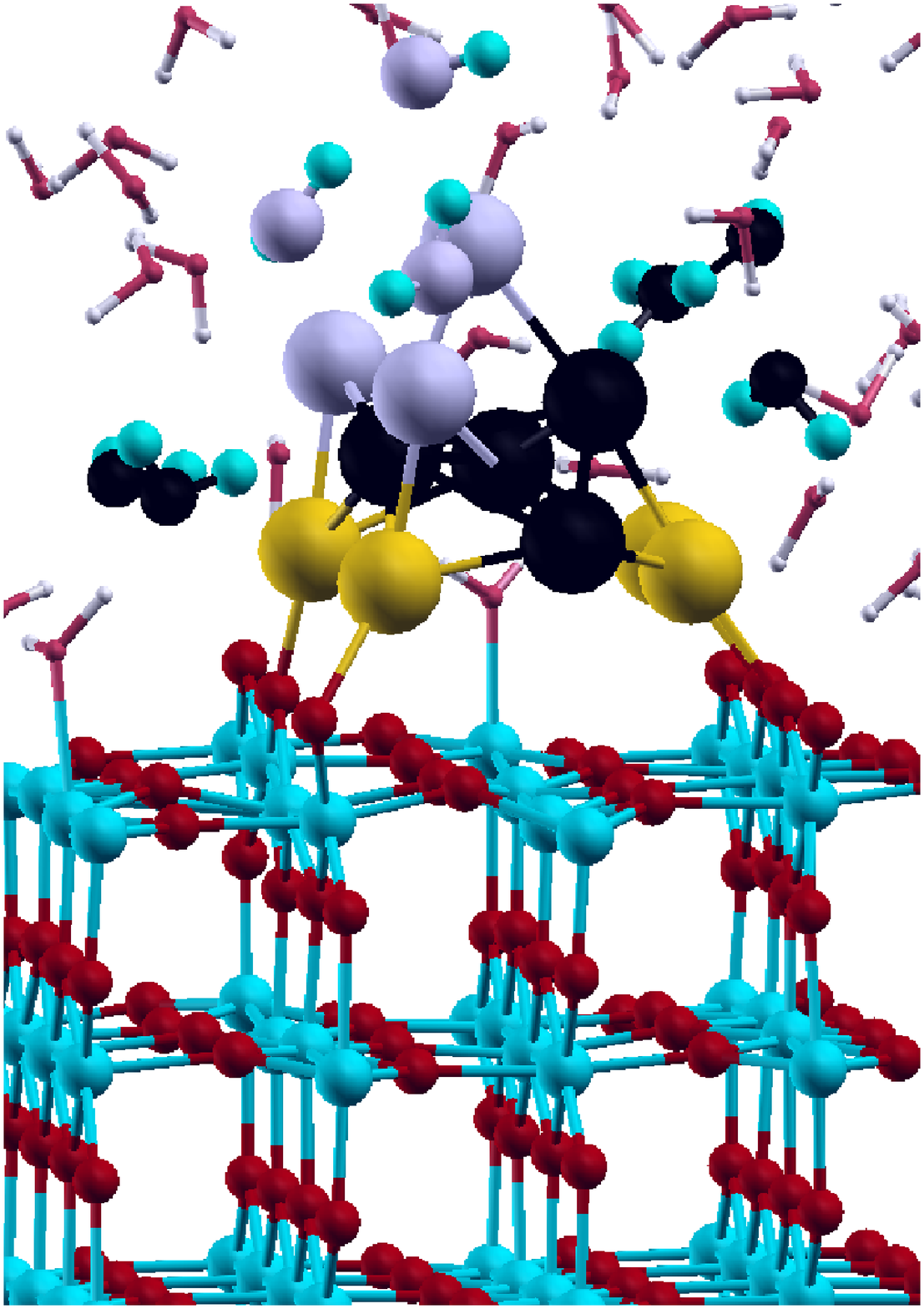}}
\subfigure[]{\includegraphics[angle=0,scale=0.37]{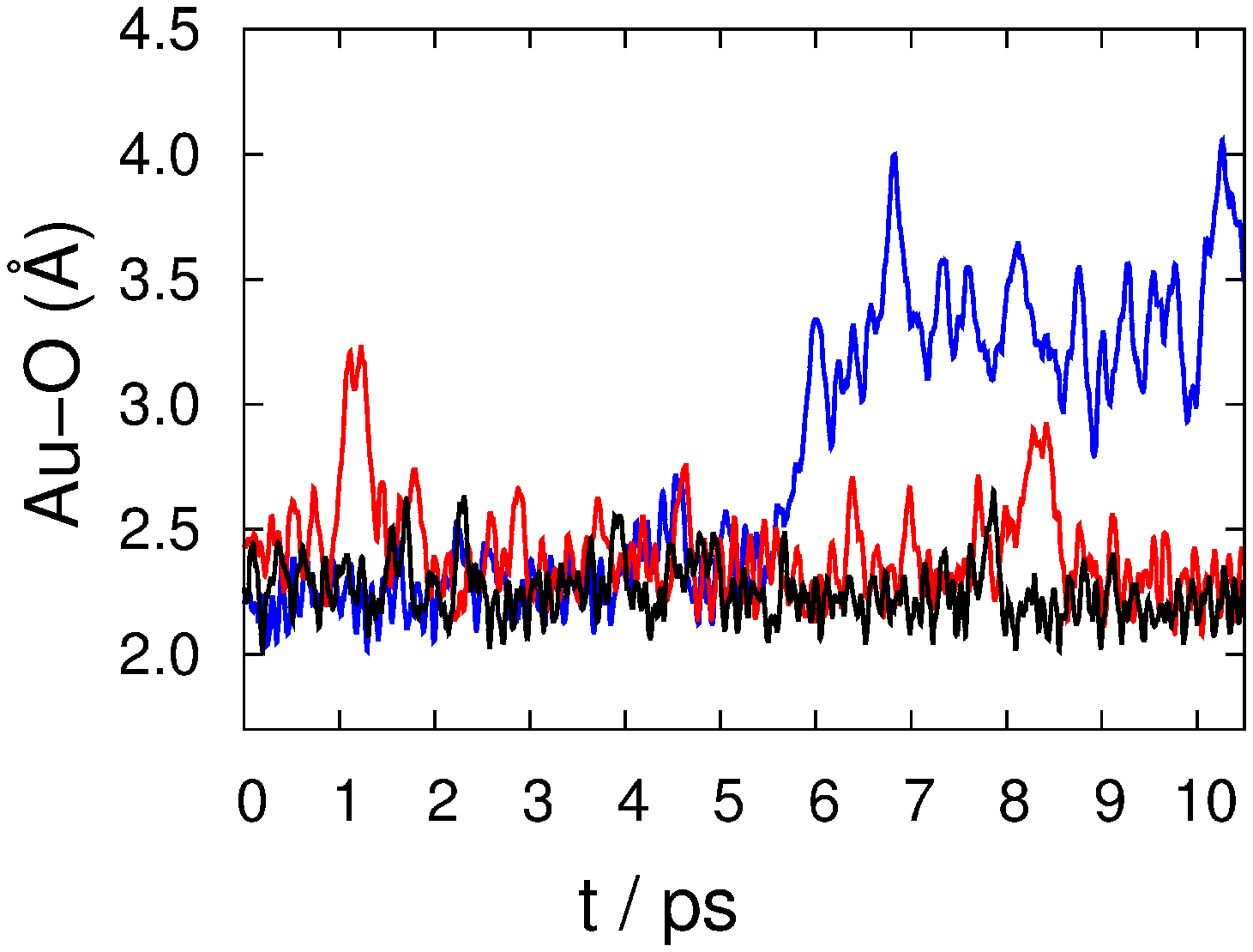}}
\end{tabular}
\caption{%
\footnotesize 
{\bf Snapshot of a typical hydration structure for the 
Au$_{11}$/TiO$_{2}$(110) nanocatalyst in aqueous solution.} 
In the cationic/anionic solvation pattern $\mathrm{H_{2}O}$ molecules 
(violet/black spheres)
are pointing with their $\mathrm{O}$/$\mathrm{H}$ atoms 
toward top/middle Au~atoms (violet/black spheres) (a).
(b) Time evolution of the
$\mathrm{Au}$--$\mathrm{O}$ distances of the gold atoms depicted in violet. 
\label{fig4}}
\end{center}
\end{figure}
As a result, top and middle $\mathrm{Au}$ sites 
experience 
opposite charge flow 
upon interaction with the solvent, as clearly
shown in Fig.~\ref{fig3}, 
which connects to a peculiar solvation pattern. 
We thus conclude by discussing the solvation pattern around the
Au$_{11}$/TiO$_{2}$(110) nanocatalyst,
see Fig.~\ref{fig4}(a), 
which is found to feature both 
so-called ``cationic'' and ``anionic'' solvation depending
on fluctuation and polarization effects. 
This is similar to
what we have observed previously for a single $\mathrm{Au^{+}}$(aq) cation and
neutral atom, $\mathrm{Au^{0}}$(aq), in aqueous bulk solution~\cite{pccp}.
In particular, the $\mathrm{Au^{+}}$ aqua ion forms a 
quasi-linear molecular structure, 
$\mathrm{H_{2}O}$-$\mathrm{Au^{+}}$-$\mathrm{OH_{2}}$, 
in which two water molecules are firmly bound through their oxygens to the
$\mathrm{Au^{+}}$ site, as expected for a cation,
yielding an $\mathrm{Au^{+}}$-$\mathrm{O}$ 
distance of $\sim 2.04$~{\AA}.
Upon adding an electron to $\mathrm{Au^{+}}$ to yield Au$^0$,
we found that its solvation pattern changes distinctly.
The latter features two distinct regions, supporting a Janus-type behavior.
In one region a typical
anionic solvation pattern prevails where two water molecules
are pointing with their $\mathrm{H}$ atoms toward the metal. 
In other regions water binds via its $\mathrm{O}$ site to the 
$\mathrm{Au^{0}}$(aq), thus giving rise to a typical 
cationic solvation pattern.

Around the nanocatalyst, two distinct solvation patterns corresponding to different
$\mathrm{Au}$ charge states can be identified.
The solvation around middle $\mathrm{Au}$ atoms consists of 
$\mathrm{H_{2}O}$ molecules pointing with their $\mathrm{H}$ atoms toward
the $\mathrm{Au}$ atoms like in an anionic solvation shell. 
In addition, several water exchange processes are observed as for
$\mathrm{Au^{0}}$(aq) in the bulk~\cite{pccp}. 
At variance with those sites, the solvation around top $\mathrm{Au}$ atoms  
features a typical cationic solvation pattern, where a single $\mathrm{H_{2}O}$
molecule point to a top $\mathrm{Au}$ atom {\it via} its $\mathrm{O}$~site.
%
%
%mf added informations on orbitals involved in the Au-O(H2) bond
%
A projected 
%DM DOS 
density of states
analysis reveals that the 
%DM bonding 
interaction
between gold atoms and 
%DM Au and O of water 
solvating water molecules
results from the overlap between the Au(d) 
and O(p) states of the 
%DM
respective 
gold and oxygen sites.
%DM bonding.}
%
Interestingly, no $\mathrm{H_{2}O}$ exchange processes are observed in
this case again in line with the $\mathrm{Au^{+}}$(aq) cation case~\cite{pccp}.
As depicted in Fig.~\ref{fig1}, the average
L\"owdin charges of top and middle
$\mathrm{Au}$ sites are in qualitative agreement to those computed for a
single gold cation (10.924~$|e|$) and
neutral gold atom (11.210~$|e|$)
in aqueous bulk solution.
Based on this comparison, 
we are in a position to explain the sudden switch of the
L\"owdin charge involving a specific top $\mathrm{Au}$ atom and 
taking place at $\sim 6$~ps in
Figs.~\ref{fig1}(b) and \ref{fig2}.    
In Fig.~\ref{fig4}(b) we plot the distances between the top $\mathrm{Au}$ atoms with
typical ``cationic'' solvation pattern and the $\mathrm{O}$ atom of its solvating
$\mathrm{H_{2}O}$ molecule. 
This demonstrates that two $\mathrm{Au}$ atoms are
always ``bonded'' to the the same $\mathrm{H_{2}O}$ molecule along 
the full trajectory, the time-averaged distances between $\mathrm{Au}$
and $\mathrm{O}$ sites being 2.25 and 2.35~$\mathrm{\AA}$. 
The situation is very different at the third top $\mathrm{Au}$~site.
This $\mathrm{Au}$ atom is initially  
cationically solvated yielding a 
$\mathrm{Au}$--$\mathrm{O}$ distance 
around 2.3~$\mathrm{\AA}$.   
At $\sim 6$~ps the $\mathrm{H_{2}O}$ 
molecule rearranges and points to the same gold atom with one of its
$\mathrm{H}$ atoms, a fluctuation which induces 
the change of charge state of this Au~site discussed earlier. 
This reveals the intimate connection between the creation
of excess charge at specific nanocluster sites and the preferential 
orientation of the interfacial water molecules. 
%
%
%mf  added text to make connection with proposed models and experiment for WGSR
%
As discussed in the introduction for the carboxyl and redox mechanisms proposed
for the WGS reaction on metal/oxide surfaces, the 
first step involves 
%dm starting point is 
H$_{2}$O molecules that are 
initially adsorbed on the metal cluster such that their O atoms
are coordinated to an Au~site.
Therefore our simulations suggest that top site Au~atoms, which feature a 
typical ``cationic'' solvation pattern at the water-gold interface, 
%dm represent
are the most promising candidates for being 
the active site for CO oxidation via the WGS reaction,
%dm ?? is this statement OK with you : 
whereas ``anionically'' coordinated Au~sites are expected to be chemically inert. 
The very same 
%dm mechanism will hold for reactants, intermediates, or products of
argument will hold for 
%dm xxxx
other liquid-phase 
catalytic reactions~-- thus providing a novel
mechanism to create active sites at catalyst--water interfaces
in much more general terms.

In conclusion, 
finite temperature $\mathrm{PBE+U}$ simulations 
have been employed to investigate solvent-metal interactions on the 
gold/titania nanocatalyst in liquid water. 
Comparing our liquid phase to gas phase and bulk solvation 
reference data demonstrates
that interfacial water alters both the structure and electronic properties
of the supported metal cluster in a significant way,
including the creation of active sites. 
The fundamental phenomena observed will have a profound impact on 
understanding the role of solvent in heterogeneously catalyzed 
liquid-phase reactions, where 
de- and re-solvation processes play a key role. 

\

{\bf Methods}
\footnotesize

\    

All calculations have been performed using spin--polarized
$\mathrm{PBE+U}$~\cite{pbe},
and ultrasoft pseudopotentials~\cite{vanderbilt}
as implemented in {\tt CPMD}~\cite{cpmd}
and {\tt Quantum} {\tt Espresso}~\cite{QE}.
In line with our previous work~\cite{cite-also-our-basic-PRB-2009,PRL,PRB_ma},
the value of $\mathrm{U}$ = 4.2~eV was adopted.
%
%mf  added information on charge dependence on U value
%
We have carefully checked the charge distribution dependence 
on $\mathrm{U}$ by
%DM plotting 
re-computing 
%DM the time evolution of 
the L\"owdin charges of the top and middle $\mathrm{Au}$ atoms of 
the supported gold nanocluster in liquid water 
%DM 
for selected snapshots along the trajectory shown in Fig.~\ref{fig1}(b) 
using four different $\mathrm{U}$ values 
and found no significant change
(see Supporting Material
%DM Supplementary Information 
for data).

The AIMD simulations~\cite{marx-hutter-book}
used Car--Parrinello propagation,
the canonical ensemble was established at 450~K
with a Nos$\mathrm{\acute{e}}$-Hoover
chain
thermostat, and
about $\sim 10$~ps trajectories have been generated for analysis
(see Supporting Material
%DM Supplementary Information 
for details).

\

{\bf Corresponding Author}
\footnotesize

\
E-mail: matteo.farnesi@theochem.rub.de

\

{\bf Acknowledgements}
\footnotesize

\
We thank Martin Muhler for fruitful discussions.
Partial financial support from 
Research Department ``Interfacial Systems Chemistry'' and 
the Cluster of Excellence RESOLV (EXC~1069) 
%dm
funded by Deutsche Forschungsgemeinschaft 
is gratefully acknowledged.
Computational resources were provided by 
NIC (J\"ulich), {\sc Bovilab@RUB} (Bochum), RV--NRW 
as well as PRACE (FERMI at Cineca).

\

{\bf Supporting Information}
%DM Supporting Information (SI)}
\footnotesize

\

Detailed descriptions of methods used for the
calculations, model system and and much additional analyses. This material is
available free of charge via the
Internet at http://pubs.acs.org.

\end{document}